\RequirePackage{fix-cm}
\documentclass[smallextended]{svjour3wide}       % onecolumn (second format)
\smartqed  % flush right qed marks, e.g. at end of proof
\usepackage{graphicx}
\journalname{Journal of Low Temperature Physics}

\makeatletter
\def\Left#1#2\Right{\begingroup%
   \def\ts@r{\nulldelimiterspace=0pt \mathsurround=0pt}%
   \let\@hat=#1%
   \def\sht@im{#2}%
   \def\@t{{\mathchoice{\def\@fen{\displaystyle}\k@fel}%
          {\def\@fen{\textstyle}\k@fel}%
          {\def\@fen{\scriptstyle}\k@fel}%
          {\def\@fen{\scriptscriptstyle}\k@fel}}}%
   \def\g@rin{\ts@r\left\@hat\vphantom{\sht@im}\right.}%
   \def\k@fel{\setbox0=\hbox{$\@fen\g@rin$}\hbox{%
      $\@fen \kern.3875\wd0 \copy0 \kern-.3875\wd0%
      \llap{\copy0}\kern.3875\wd0$}}%
      \def\pt@h{\mathopen\@t}\pt@h\sht@im%
      \Right}%
\def\Right#1{\let\@hat=#1%
   \def\st@m{\mathclose\@t}%
   \st@m\endgroup}
\makeatother
\begin{document}

\title{Bogoliubov Theory for a Superfluid Bose Gas Flowing in a Random Potential: Stability and Critical Velocity}

\titlerunning{Bogoliubov Theory for a Bose Gas Flowing in a Random Potential} % if too long for running head

\author{Taiki Haga}


\institute{T. Haga \at
              Department of Physics, Kyoto University, Kyoto 606-8502, Japan \\
              \email{haga@scphys.kyoto-u.ac.jp}           %  \\
}

\date{Received: date / Accepted: date}

\maketitle

\begin{abstract}
We investigate the stability and critical velocity of a weakly interacting Bose gas flowing in a random potential.
By applying the Bogoliubov theory to a disordered Bose system with a steady flow, the condensate density and the superfluid density are determined as functions of the disorder strength, flow velocity, and temperature.
The critical velocity, at which the steady flow becomes unstable, is calculated from the spectrum of hydrodynamic excitation.
We also show that in two dimensions the critical velocity strongly depends on the system size.
\end{abstract}

\section{Introduction}
\label{intro}

% The breakdown of superfluidity %
The superfluidity, which is one of the most remarkable properties of interacting Bose systems, is characterized by the manifestation of a nondissipative flow.
When its flow velocity exceeds a critical velocity $V_{\rm c}$, the superfluidity is broken and dissipation sets in. 
The breakdown of superfluidity has been observed in experiments with liquid $^{4}{\rm He}$ \cite{Allum} and cold atomic gases 
\cite{Chikkatur,Inouye,Sarlo,Engels,Mun,Mckay,Neely} in the presence of a defect potential or an optical lattice.
These experiments imply that the nucleation of solitons or vortices is responsible for the breakdown of superfluidity.

% Landau and Dynamical instability %
The stability of flowing Bose systems has been studied theoretically by using the Gross-Pitaevskii (GP) equation \cite{Frisch,Hakim,Winiecki,Huepe,Pavloff,Danshita07,Sasaki,Kato,Dubessy,Kunimi}.
One can distinguish two types of stability \cite{Machholm,Wu,Iigaya}.
The first is Landau stability, and the criterion for this is that a change in the energy functional with respect to a small perturbation in the condensate wave function always be positive.
If the system is Landau unstable, it can lower its energy by emitting phonon-like excitations to a thermal cloud surrounding the condensate.
In a homogeneous environment, the critical velocity corresponding to the Landau instability is equal to the velocity of sound \cite{Wu}.
The second type is dynamical stability, and the criterion for this is that the linearized time-dependent GP equation has no complex eigenvalues.
If the system is dynamically unstable, a small fluctuation around the original state grows exponentially in time and, eventually, solitons or vortices are nucleated.
In other words, Landau instability is caused by the transfer of energy from the condensate to the thermal excitations, while dynamical instability is caused by the collective modulation of the condensate.
Thus, in the low temperature regime, dynamical instability plays a dominant role in the breakdown of a superfluid \cite{Sarlo}.
It is worth noting that the critical velocity corresponding to Landau instability is lower than that corresponding to dynamical instability.

% Superfluid in disordered environment %
In this paper, we consider the Landau and dynamical stability of a superfluid Bose gas flowing in a random potential.
Recently, the transport property of a superfluid Bose gas in a disordered environment has attracted considerable attention.
The decay of supercurrents in the presence of disorder is experimentally observed by studying the damping of dipole oscillations in quasi-one-dimensional systems \cite{Dries,Tanzi}.
The stability of superfluid is investigated theoretically for one-dimensional disordered Bose gases by using the GP equation in Refs.~\cite{Paul} and \cite{Albert}.
Although in one dimension supercurrents decay at any flow velocity due to phase slip induced by quantum fluctuations \cite{Polkovnikov,Danshita13}, the qualitative behavior of a velocity-dependent damping coefficient predicted from the GP equation is similar to that observed in experiments.
Unfortunately, these studies in Refs.~\cite{Paul} and \cite{Albert} cannot be extended to higher-dimensional systems, because they rely on a peculiarity of the one-dimensional GP equation.
Furthermore, the GP equation cannot take into account the effect of thermal excitation at finite temperature.

% The purpose of this paper %
The purpose of this study is to develop a general approach for treating a Bose gas flowing in a random potential and to determine the critical velocity for weak and moderate disorder.
We consider the Bose-Hubbard model with a random potential that obeys a Gaussian distribution.
By applying the Bogoliubov theory to the disordered Bose-Hubbard model with a steady flow, we determine the condensate density and the superfluid density as functions of the strength of the random potential, flow velocity, and temperature.
We discuss the Landau and dynamical stability of the steady flow by investigating the spectrum of hydrodynamic collective excitation.
The critical velocity is determined as the flow velocity at which the steady flow becomes unstable.
It is a decreasing function with respect to the disorder strength and temperature.
Furthermore, the system size dependence of the critical velocity is investigated in two dimensions, and we find that it decreases as the system size increases.

% Outline of the paper %
This paper is organized as follows:
In Sec.~\ref{sec:Model}, we introduce the Bose-Hubbard model with a random potential and summarize its phase structure in equilibrium.
In Sec.~\ref{sec:Bogoliubov_theory_with_disorder}, we review the Bogoliubov theory for a disordered Bose gas in equilibrium.
We emphasize that there are two types of formulations depending on how to decompose the field operator. 
In Sec.~\ref{sec:General_formalism}, we develop the Bogoliubov theory for the disordered Bose-Hubbard model in the presence of a flow.
The condensate density and the momentum distribution function of the noncondensed particles are expressed as functions of the strength of the random potential, flow velocity, and temperature.
We also discuss the conditions for Landau and dynamical stability.
In Sec.~\ref{sec:Results}, we calculate the critical velocity as a function of the strength of the random potential and temperature in two dimensions.
The system size dependence of the critical velocity is also discussed.
In Sec.~\ref{sec:Summary}, we summarize our results.

\section{Model}
\label{sec:Model}

We consider the Bose-Hubbard model with a random potential:
\begin{equation}
\hat{H}=-J \sum_{\langle ij \rangle} (\hat{b}^{\dag}_i \hat{b}_j+\hat{b}^{\dag}_j \hat{b}_i) +\frac{U}{2} \sum_i \hat{b}^{\dag}_i \hat{b}^{\dag}_i \hat{b}_i \hat{b}_i + \sum_i V_i \hat{b}^{\dag}_i \hat{b}_i,
\label{Hamiltonian}
\end{equation}
where $\hat{b}_i$ denotes the annihilation operator at site $i$, $ \langle ij \rangle $ indicates the nearest-neighbor sites on a $d$-dimensional hypercubic lattice, and the positive parameters $J$ and $U$ are the hopping energy and the interaction energy, respectively.
The random potential $V_i$ obeys a Gaussian distribution with
\begin{equation}
\Left< V_i \Right> =0, \:\:\:  \Left< V_i V_j \Right> =V_0^2 \delta_{ij},
\end{equation}
where $ \Left< ... \Right> $ denotes the average with respect to the random potential.
The number operator at site $i$ is defined by $\hat{n}_i=\hat{b}^{\dag}_i \hat{b}_i$.
We also denote $n$ as the mean number density $ n=N_s^{-1} \sum_i \langle \hat{n}_i \rangle $, where $N_s=L^d$ is the number of sites, and $ \langle ... \rangle $ represents the total average,
\begin{equation}
\langle \hat{A} \rangle = \Left< \frac{\mathrm{Tr} \hat{A} e^{-\beta \hat{H}} }{\mathrm{Tr} e^{-\beta \hat{H}}} \Right>.
\end{equation}
For later convenience, we also introduce the notation of the thermal average,
\begin{equation}
\langle \hat{A} \rangle_{H} = \frac{\mathrm{Tr} \hat{A} e^{-\beta \hat{H}} }{\mathrm{Tr} e^{-\beta \hat{H}}},
\end{equation}
which is a function of the random potential.
An advantage of lattice systems is that they are free from the ambiguities associated with ultraviolet divergences that are found in continuous systems with contact interactions.

The phase structure of the disordered Bose-Hubbard model has been extensively investigated theoretically 
\cite{Fisher,Singh-92,Freericks,Buonsante,Bissbort,Altman,Soyler,Ristivojevic,Zuniga} and experimentally \cite{White,Deissler,D'Errico}.
It has three distinct phases: the superfluid phase, the Bose glass phase, and the Mott phase.
For small $U/J$ and $V_0/J$, the system is a superfluid.
When $V_0/J$ is sufficiently large, the superfluid phase is broken and the Bose glass phase appears.
This insulating phase is characterized by a finite compressibility and the absence of a gap in the one-particle spectrum.
When the distribution of the random potential is bounded and the mean number density $n$ is an integer, for large $U/J$ and $U/V_0$ the system becomes a Mott insulator.
Note that for the Gaussian distributed random potential, there does not exist any gapped insulating phase because the presence of the unbounded disorder inevitably closes the energy gap.
In our study, we consider a weakly interacting system satisfying $Un/J \ll 1$ for which the Bogoliubov theory is valid.
This condition also means that the healing length $ \sim \sqrt{J/(Un)} $ is much larger than the correlation length of the random potential.

\section{Bogoliubov theory with disorder}
\label{sec:Bogoliubov_theory_with_disorder}

In this section, we review the Bogoliubov theory for a disordered Bose gas in equilibrium.
The main ideas of the Bogoliubov theory are (i) to decompose the field operator $\hat{b}_i$ into a condensate wave function and a fluctuation around it, (ii) to expand the Hamiltonian up to quadratic order of the fluctuation, (iii) to diagonalize it by using the Bogoliubov transformation \cite{Fetter}.
For homogeneous cases, this decomposition into a uniform condensate and a quantum fluctuation $\hat{b}_i=\sqrt{n_0}+\hat{\psi}_i$ is unique and well-defined.
In contrast, in the presence of the random potential, an ambiguity arises in the decomposition because there are two types of ``fluctuations''.
To clarify this subtle point, we formally decompose the field operator into a uniform background, the deformation of the condensate due to the random potential, and the quantum fluctuation,
\begin{equation}
\hat{b}_i = \sqrt{n_0} + \hat{\psi}_{\mathrm{R},i} + \hat{\psi}_{\mathrm{Q},i},
\label{decomposition_R_Q}
\end{equation}
where two fluctuations $\hat{\psi}_{\mathrm{R},i}$ and $\hat{\psi}_{\mathrm{Q},i}$ satisfy
\begin{eqnarray}
\langle \hat{\psi}_{\mathrm{Q},i} \rangle_H = 0, \nonumber \\
\langle \hat{\psi}_{\mathrm{R},i} \rangle_H \neq 0, \:\:\: \langle \hat{\psi}_{\mathrm{R},i} \rangle = 0.
\label{psi_R_Q_condition}
\end{eqnarray}
Thus, we have two types of decomposition schemes depending on whether $\hat{\psi}_{\mathrm{R},i}$ is considered as background or fluctuation.

In the first scheme, we decompose $\hat{b}_i$ into the deformed condensate and the quantum fluctuation,
\begin{equation}
\hat{b}_i = \phi_i + \hat{\psi}_{1,i},
\label{decomposition_1}
\end{equation}
where $\phi_i=\sqrt{n_0} + \hat{\psi}_{\mathrm{R},i}$ and $\hat{\psi}_{1,i}=\hat{\psi}_{\mathrm{Q},i}$.
In equilibrium, $\phi_i$ can be assumed to be real without loss of generality.
The condition Eq.~(\ref{psi_R_Q_condition}) implies that the condensate wave function $\phi_i$ satisfies the discretized GP equation,
\begin{eqnarray}
-J \sum_{\langle j \rangle_i} \phi_j + U \phi_i^3 + V_i \phi_i = \mu \phi_i,
\end{eqnarray}
where we have introduced the chemical potential $\mu$.
The notation $\langle j \rangle_i$ means the nearest-neighbor sites of $i$.
By substituting Eq.~(\ref{decomposition_1}) into the original Hamiltonian, in the leading order, we have 
\begin{eqnarray}
H_1 = -J \sum_{\langle ij \rangle} (\hat{\psi}^{\dag}_{1,i} \hat{\psi}_{1,j}+\hat{\psi}^{\dag}_{1,j} \hat{\psi}_{1,i}) + \sum_i (V_i-\mu) \hat{\psi}^{\dag}_{1,i} \hat{\psi}_{1,i} \nonumber \\
+ \sum_i U \phi_i^2 \Bigl( 2 \hat{\psi}^{\dag}_{1,i} \hat{\psi}_{1,i} + \frac{1}{2} \hat{\psi}^{\dag}_{1,i} \hat{\psi}^{\dag}_{1,i} + \frac{1}{2} \hat{\psi}_{1,i} \hat{\psi}_{1,i} \Bigr).
\label{Hamiltonian_1}
\end{eqnarray}
By diagonalizing this Hamiltonian, we obtain the excitation spectrum for quasi-particles.

In the second scheme, we decompose $\hat{b}_i$ into a uniform background and a fluctuation around it,
\begin{equation}
\hat{b}_i = \sqrt{n_0} + \hat{\psi}_{2,i},
\label{decomposition_2}
\end{equation}
where $\hat{\psi}_{2,i}=\hat{\psi}_{\mathrm{R},i} + \hat{\psi}_{\mathrm{Q},i}$.
From the condition Eq.~(\ref{psi_R_Q_condition}), the uniform background satisfies 
\begin{equation}
(-zJ+U n_0 - \mu) \sqrt{n_0} =0,
\end{equation}
where $z$ is the coordination number of the lattice.
By substituting Eq.~(\ref{decomposition_2}) into the original Hamiltonian, in the leading order, we have 
\begin{eqnarray}
H_2 = -J \sum_{\langle ij \rangle} (\hat{\psi}^{\dag}_{2,i} \hat{\psi}_{2,j}+\hat{\psi}^{\dag}_{2,j} \hat{\psi}_{2,i}) + \sum_i \sqrt{n_0} V_i (\hat{\psi}^{\dag}_{2,i} + \hat{\psi}_{2,i})  \nonumber \\
+ \sum_i U n_0 \Bigl( 2 \hat{\psi}^{\dag}_{2,i} \hat{\psi}_{2,i} + \frac{1}{2} \hat{\psi}^{\dag}_{2,i} \hat{\psi}^{\dag}_{2,i} + \frac{1}{2} \hat{\psi}_{2,i} \hat{\psi}_{2,i} \Bigr) - \sum_i \mu \hat{\psi}^{\dag}_{2,i} \hat{\psi}_{2,i},
\label{Hamiltonian_2}
\end{eqnarray}
where the term $V_i \hat{\psi}^{\dag}_{2,i} \hat{\psi}_{2,i}$ has been omitted because it is negligible compared to the term $ \sqrt{n_0} V_i \hat{\psi}_{2,i}$.

Unfortunately, in general, these two decomposition schemes yield different results.
The reason is that the formal decomposition Eq.~(\ref{decomposition_R_Q}) is not well-defined.
Since the effects of the disorder and the quantum fluctuation are mutually coupled in a complicated way, one cannot split them in terms of the decomposition of the field operator.
For example, one can check that, in the lowest-order of the strength of the interaction and the disorder, these two decomposition schemes yield same deformed condensate wave functions $\phi_i$ and $\sqrt{n_0} + \langle \hat{\psi}_{2,i} \rangle_H$, but different energy densities.
To the best of our knowledge, there is no systematic study making a detailed comparison between these schemes.

The first decomposition scheme is expected to be preferable when the random potential varies slowly in space; the correlation length of the random potential is larger than the healing length of the condensate.
The reason is that in such cases the condensate can be readily deformed according to a given configuration of the random potential.
In addition, the first decomposition scheme is also useful to investigate the short length scale structure of the quasi-particle excitation modes \cite{Singh-94,Gaul,Lugan,Fontanesi,Saliba}.
The short length scale means the length scale comparable to the correlation length of the random potential.
The disadvantage of this scheme is its analytical intractability.
In fact, since Eq.~(\ref{Hamiltonian_1}) cannot be diagonalized analytically, we have to employ a computationally demanding numerical method, which limits the system sizes we can consider.
Furthermore, it is hard to take into account the effect of the higher order terms within the framework of this scheme.

The second decomposition scheme has been often employed in early studies because Eq.~(\ref{Hamiltonian_2}) can be easily diagonalized by the standard Bogoliubov transformation.
This approach was first applied to a weakly interacting disordered Bose gas by Huang and Meng \cite{Huang}, who considered the cases of asymptotically weak disorder.
The advantage of this decomposition scheme is that it can be extended to the case of finite temperature, interaction, and disorder by taking into account the effects of the higher order terms in a systematic way \cite{Falco,Yukalov07-1,Yukalov07-2}.
Considering this analytical tractability, we employ this decomposition scheme in this study.
We expect that, for the short-range random potential whose correlation length is much smaller than the healing length of the condensate, this scheme yields reasonable results for the thermodynamic quantities and the long-wavelength excitation spectrum because the local structure of the random potential may not have major influence on the large-scale behavior.

Finally, note that there is also an ambiguity in the definition of ``condensed'' and ``noncondensed'' particles.
In the first decomposition scheme, particles scattered from the zero-momentum state by the random potential are included in ``condensed'' particles, while in the second decomposition scheme, they are considered as ``noncondensed'' particles.

\section{General formalism}
\label{sec:General_formalism}

\subsection{Failure of the Landau criterion}

In this study, we calculate the critical velocity of the disordered Bose-Hubbard model.
According to the usual Landau criterion, the critical velocity is identical to the sound velocity at equilibrium.
Thus, one may consider that the main problem is simply to calculate the sound velocity as a function of the disorder.
However, this approach can lead to qualitatively incorrect results even if the disorder is weak.
In what follows, we discuss why the usual Landau criterion is not appropriate for evaluating the critical velocity in the presence of the disorder.

For an interacting Bose gas at rest, the spectrum of excitation is given by $\Omega_{\mathrm{rest}}(\mathbf{q}) = c_0 |\mathbf{q}|$ near $\mathbf{q} = \bf{0}$, where $c_0$ is the sound velocity at equilibrium.
In a homogeneous environment, the spectrum of excitation for the system flowing at a velocity $\mathbf{v}=v \mathbf{e_x}$ is given by 
\begin{equation}
\Omega_{\mathrm{move}}(\mathbf{q}) = \Omega_{\mathrm{rest}}(\mathbf{q}) + v q_x,
\label{Omega_move}
\end{equation}
in the moving frame.
For $v > c_0$, since $\Omega_{\mathrm{move}}(\mathbf{q})$ has negative values, the superfluid becomes Landau unstable.
This is the usual Landau criterion, which states that the critical velocity is identical to the sound velocity at equilibrium, $v_{\mathrm{c}}=c_0$.
Note that this criterion is valid only when there are infinitesimally small defects.
In the presence of a finite amount of disorder, Eq.~(\ref{Omega_move}) is incorrect because the translational symmetry of the system is broken.
Thus, the critical velocity can be different from the sound velocity at equilibrium.

\subsection{Bogoliubov shift to the field operator}

In the following, we determine the critical velocity by considering the stability of a steady flow.
In order to calculate the spectrum of excitation $\Omega_{\mathrm{move}}(\mathbf{q})$ directly, we extend the disordered Bogoliubov theory to situations in which a flow is present.
Since the Bogoliubov theory is invalid when the condensate fraction $n_0/n$ is small, we restrict our attention to the weak and moderate disorder regime.

According to the second decomposition scheme mentioned in the previous section, we decompose the field operator into a uniform background with a momentum $\mathbf{K}$ and a fluctuation around it,
\begin{equation}
\hat{b}_j=(\sqrt{n_0}+\hat{\psi}_j)e^{i \mathbf{K} \cdot \mathbf{R_j}},
\label{Bogoliubov_shift}
\end{equation}
where $n_0$ denotes the density of the condensed particles, $\mathbf{R_j}$ is the position of site $j$,
$ \sqrt{n_0} e^{i \mathbf{K} \cdot \mathbf{R_j}} $ is the wave function of the condensed particles, and 
$\hat{\psi}_j$ is the operator of the noncondensed particles in the moving frame.
The condensate wave function and the operator $\hat{\psi}_j$ are taken to be mutually orthogonal:
\begin{equation}
\sum_i \hat{\psi}_i =0.
\end{equation}
In the following, the flow is assumed to be parallel to the $x$-direction, $\mathbf{K}=K \mathbf{e_x}$.

Substituting Eq.~(\ref{Bogoliubov_shift}) into the Hamiltonian (\ref{Hamiltonian}), we have
\begin{equation}
\hat{H}=\hat{H}^{(0)}+\hat{H}^{(2)}+\hat{H}^{(3)}+\hat{H}^{(4)}+\hat{H}_{\rm ext},
\end{equation}
where $ \hat{H}^{(n)}$ are the terms containing the products of $n$ operators. 
By introducing $ \hat{\psi}_{\mathbf{q}} = N_s^{-1/2} \sum_j \hat{\psi}_j e^{-i \mathbf{q} \cdot \mathbf{R_j}} $, the quadratic term is given by
\begin{equation}
\hat{H}^{(2)}=\sum_{\mathbf{q} \neq \mathbf{0}} (\epsilon_{\mathbf{K+q}} + 2n_0 U ) \hat{\psi}^{\dag}_{\mathbf{q}} \hat{\psi}_{\mathbf{q}} + \frac{1}{2} n_0 U \sum_{\mathbf{q} \neq \mathbf{0}} (\hat{\psi}^{\dag}_{\mathbf{q}} \hat{\psi}^{\dag}_{\mathbf{-q}} + \hat{\psi}_{\mathbf{q}} \hat{\psi}_{\mathbf{-q}}),
\end{equation}
and the quartic term is given by
\begin{equation}
\hat{H}^{(4)}=\frac{U}{2N_s} {\sum_{\mathbf{k,p,q}}}' \hat{\psi}^{\dag}_{\mathbf{k}} \hat{\psi}^{\dag}_{\mathbf{p}} \hat{\psi}_{\mathbf{k-q}} \hat{\psi}_{\mathbf{p+q}},
\end{equation}
where the prime on the summation sign means that $\mathbf{k} \neq \mathbf{0}$, $\mathbf{p} \neq \mathbf{0}$, $\mathbf{k-q} \neq \mathbf{0}$, and $\mathbf{p+q} \neq \mathbf{0}$.
The wave number takes the values $ q_{\mu} = 2 \pi m_{\mu}/L, \: m_{\mu}=-L/2,...,L/2-1$, where $L$ is system size.
$\epsilon_{\mathbf{q}} \equiv -2J \sum_{\mu} {\rm cos}q_{\mu}$ is the kinetic energy of a free particle, and
$ \hat{H}_{\rm ext} $ denotes the contribution from the random potential,
\begin{equation}
\hat{H}_{\rm ext}=\sqrt{n_0} \sum_{\mathbf{q} \neq \mathbf{0}} (V_{\mathbf{q}} \hat{\psi}^{\dag}_{\mathbf{q}} + V_{\mathbf{q}}^* \hat{\psi}_{\mathbf{q}}) + \frac{1}{\sqrt{N_s}} \sum_{\mathbf{k,p} \neq \mathbf{0}} V_{\mathbf{k-p}} \hat{\psi}^{\dag}_{\mathbf{k}} \hat{\psi}_{\mathbf{p}}.
\label{H_ext}
\end{equation}

\subsection{Hartree-Fock-Bogoliubov approximation}

The fraction of the noncondensed particles is not small, due to the presence of the random potential.
Thus, we cannot ignore $\hat{H}^{(4)}$, which represents the interaction between the noncondensed particles, in spite of the small interaction energy $U/J \ll 1$.

We simplify the nonlinear terms $ \hat{H}^{(3)}$ and $ \hat{H}^{(4)}$ by means of the Hartree-Fock-Bogoliubov (HFB) approximation \cite{Yukalov07-1}.
First, we ignore the third-order term; $\hat{H}^{(3)} = 0$.
In order to express the result for the term $ \hat{H}^{(4)}$ in a compact form,
we introduce the momentum distribution of the noncondensed particles
\begin{equation}
\rho(\mathbf{q}) = \langle \hat{\psi}^{\dag}_{\mathbf{q}} \hat{\psi}_{\mathbf{q}} \rangle,
\end{equation}
and the momentum distribution of paired particles
\begin{equation}
\sigma(\mathbf{q}) = \langle \hat{\psi}_{\mathbf{q}} \hat{\psi}_{\mathbf{-q}} \rangle.
\end{equation}
The noncondensate density and the anomalous density are given by
\begin{equation}
n_1 = \frac{1}{N_s} \sum_{\mathbf{q} \neq \mathbf{0}} \rho(\mathbf{q}), \:\:\: \sigma_1 = \frac{1}{N_s} \sum_{\mathbf{q} \neq \mathbf{0}} \sigma(\mathbf{q}), 
\end{equation}
respectively, and $n=n_0+n_1$.
In the HFB approximation, $ \hat{H}^{(4)}$ is rewritten as
\begin{eqnarray}
\hat{H}^{(4)}= -\frac{1}{2}U(2 n_1^2+\sigma_1^2)N_s + \sum_{\mathbf{q} \neq \mathbf{0}} 2 n_1 U \hat{\psi}^{\dag}_{\mathbf{q}} \hat{\psi}_{\mathbf{q}} + \frac{1}{2} \sigma_1 U \sum_{\mathbf{q} \neq \mathbf{0}} (\hat{\psi}^{\dag}_{\mathbf{q}} \hat{\psi}^{\dag}_{\mathbf{-q}} + \hat{\psi}_{\mathbf{q}} \hat{\psi}_{\mathbf{-q}}).
\end{eqnarray}

While the HFB approximation is simple, it suffers from a notorious problem related to the existence of a gap in the one-particle excitation spectrum \cite{Griffin}, which is forbidden in the superfluid regime, according to the Hugenholtz-Pines theorem \cite{Fetter}.
To avoid this problem, we employ the representative statistical ensemble approach \cite{Yukalov06,Yukalov08}.
In this approach, we introduce two chemical potentials, $\mu_0$ and $\mu_1$, which are associated with $n_0$ and $n_1$, respectively.
The grand Hamiltonian is defined as
\begin{equation}
\hat{H}'=\hat{H}-\mu_0 n_0 N_s-\mu_1 \sum_i \hat{\psi}^{\dag}_i \hat{\psi}_i.
\label{grand_Hamiltonian}
\end{equation}
The grand thermodynamic potential is given by
\begin{equation}
\Omega = -T \Left< {\rm ln} {\rm Tr} e^{-\beta \hat{H}'} \Right>.
\end{equation}
The chemical potential $\mu_0$, which controls the number of condensed particles, is determined from 
\begin{equation}
\frac{\partial \Omega}{\partial n_0} =0.
\label{mu_0_def}
\end{equation}
The other chemical potential, $\mu_1$, which controls the number of noncondensed particles, is chosen such that the one-particle spectrum is gapless.

Thus, Eq.~(\ref{grand_Hamiltonian}) can be rewritten as follows:
\begin{equation}
\hat{H}'=E_{\rm HFB} + \sum_{\mathbf{q} \neq \mathbf{0}} \omega_{\mathbf{q}} \hat{\psi}^{\dag}_{\mathbf{q}} \hat{\psi}_{\mathbf{q}} + \frac{1}{2} \Delta \sum_{\mathbf{q} \neq \mathbf{0}} (\hat{\psi}^{\dag}_{\mathbf{q}} \hat{\psi}^{\dag}_{\mathbf{-q}} + \hat{\psi}_{\mathbf{q}} \hat{\psi}_{\mathbf{-q}}) + \hat{H}_{\rm ext},
\end{equation}
where $\omega_{\mathbf{q}}$ and $\Delta$ are defined by
\begin{equation}
\omega_{\mathbf{q}} = \epsilon_{\mathbf{K+q}}+2nU-\mu_1,
\label{omega_def}
\end{equation}
and
\begin{equation}
\Delta = (n_0+\sigma_1)U,
\label{Delta_def}
\end{equation}
respectively.
$E_{\rm HFB}$ is given by
\begin{equation}
E_{\rm HFB} = \left(\frac{1}{2}n_0 U + \epsilon_{\mathbf{K}} - \mu_0 \right) n_0 N_s -\frac{1}{2}U (2 n_1^2 + \sigma_1^2) N_s.
\end{equation}

\subsection{Stochastic mean-field approximation}

Next, we reorganize Eq.~(\ref{H_ext}), which describes the interaction of particles with the external random potential.
The linear terms in Eq.~(\ref{H_ext}) correspond to the scattering processes between condensed and noncondensed particles, due to the random potential.
The nonlinear terms correspond to scattering processes among noncondensed particles.
For weak disorder, one can omit the nonlinear terms, as was done by Huang and Meng \cite{Huang}.
Since we consider a situation with moderate disorder, we must retain these terms, and so we use the stochastic mean-field approximation \cite{Yukalov07-1}.

In the stochastic mean-field approximation, we define
\begin{equation}
\alpha_i = \langle \hat{\psi}_i \rangle_{H},
\label{alpha_def}
\end{equation}
where $ \langle ... \rangle_{H} $ denotes the thermal average for a fixed random potential $V_i$.
Note $\alpha_i$ is a function of $V_i$ and it satisfies $ \Left< \alpha_i \Right> = 0 $.
Using this $\alpha_i$, each term in the second summation in Eq.~(\ref{H_ext}) can be approximated as
\begin{equation}
V_{\mathbf{k-p}} \hat{\psi}^{\dag}_{\mathbf{k}} \hat{\psi}_{\mathbf{p}} \simeq V_{\mathbf{k-p}} (\alpha_{\mathbf{k}}^* \hat{\psi}_{\mathbf{p}} + \hat{\psi}^{\dag}_{\mathbf{k}} \alpha_{\mathbf{p}} - \alpha_{\mathbf{k}}^* \alpha_{\mathbf{p}}).
\end{equation}
We introduce the effective random potential as 
\begin{equation}
\varphi_{\mathbf{q}} = \sqrt{n_0} V_{\mathbf{q}} + \frac{1}{\sqrt{N_s}} \sum_{\mathbf{q'} \neq \mathbf{0}} \alpha_{\mathbf{q'}} V_{\mathbf{q-q'}},
\label{phi_def}
\end{equation}
and then Eq.~(\ref{H_ext}) can be rewritten as
\begin{equation}
\hat{H}_{\rm ext} = E_{\rm ext} + \sum_{\mathbf{q} \neq \mathbf{0}} (\varphi_{\mathbf{q}} \hat{\psi}^{\dag}_{\mathbf{q}} + \varphi_{\mathbf{q}}^* \hat{\psi}_{\mathbf{q}}),
\end{equation}
where $E_{\rm ext}$ is given by
\begin{equation}
E_{\rm ext} = -\frac{1}{\sqrt{N_s}} \sum_{\mathbf{p,q} \neq \mathbf{0}} \alpha_{\mathbf{p}}^* \alpha_{\mathbf{q}} V_{\mathbf{p-q}}.
\end{equation}

\subsection{Self-consistent equations}
\label{sec:Self-consistent equations}

We employ the Bogoliubov transformation
\begin{equation}
\hat{\psi}_{\mathbf{q}} = u_{\mathbf{q}} \hat{c}_{\mathbf{q}} + v_{\mathbf{q}} \hat{c}^{\dag}_{\mathbf{-q}} + \alpha_\mathbf{q},
\label{Bogoliubov_transformation}
\end{equation}
where $u_{\mathbf{q}},\:v_{\mathbf{q}}$ are assumed to be real and satisfy $ u_{\mathbf{q}} = u_{\mathbf{-q}},\:v_{\mathbf{q}} = v_{\mathbf{-q}} $.
Substituting these into the Hamiltonian, $u_{\mathbf{q}}$, $v_{\mathbf{q}}$, and $\alpha_{\mathbf{q}}$ are determined such that the off-diagonal terms and linear terms vanish.
Finally, we have
\begin{equation}
\hat{H'} = E_{\rm B} + \sum_{\mathbf{q} \neq \mathbf{0}} E_{\mathbf{q}} \hat{c}^{\dag}_{\mathbf{q}} \hat{c}_{\mathbf{q}},
\end{equation}
and $u_{\mathbf{q}}$, $v_{\mathbf{q}}$ are calculated as
\begin{eqnarray}
u_{\mathbf{q}} = \left( \frac{1}{2} \left[ \frac{\frac{1}{2} (\omega_{\mathbf{q}}+\omega_{\mathbf{-q}})}{\{\frac{1}{4}(\omega_{\mathbf{q}}+\omega_{\mathbf{-q}})^2 -\Delta^2 \}^{1/2}} + 1 \right] \right)^{1/2}, \nonumber \\
v_{\mathbf{q}} = -\left( \frac{1}{2} \left[ \frac{\frac{1}{2} (\omega_{\mathbf{q}}+\omega_{\mathbf{-q}})}{\{\frac{1}{4}(\omega_{\mathbf{q}}+\omega_{\mathbf{-q}})^2 -\Delta^2 \}^{1/2}} - 1 \right] \right)^{1/2}.
\end{eqnarray}
The spectrum $E_{\mathbf{q}}$ is given by
\begin{equation}
E_{\mathbf{q}} = \frac{1}{2} (\omega_{\mathbf{q}}-\omega_{\mathbf{-q}}) + \left\{ \frac{1}{4}(\omega_{\mathbf{q}}+\omega_{\mathbf{-q}})^2 - \Delta^2 \right\}^{1/2}.
\label{E_q}
\end{equation}
$E_{\rm B}$ is written as
\begin{equation}
E_{\rm B}=E_{\rm HFB}+E_{\rm ext}+\frac{1}{2} \sum_{\mathbf{q} \neq \mathbf{0}} (E_{\mathbf{q}} - \omega_{\mathbf{q}}) - \sum_{\mathbf{q} \neq \mathbf{0}} \frac{1}{E_{\mathbf{q}}} \Left< |\varphi_{\mathbf{q}} u_{\mathbf{q}} + \varphi_{\mathbf{-q}}^* v_{\mathbf{q}}|^2 \Right>.
\end{equation}
By requiring that the spectrum be gapless, $ \lim_{\mathbf{q} \to \mathbf{0}} E_{\mathbf{q}} = 0 $, the chemical potential of the noncondensed particles is determined as 
\begin{equation}
\mu_1 = \epsilon_{K} + (n+n_1-\sigma_1)U.
\end{equation}
Thus, Eq.~(\ref{omega_def}) can be reduced to
\begin{equation}
\omega_{\mathbf{q}} = \epsilon_{\mathbf{K+q}}-\epsilon_{\mathbf{K}}+\Delta.
\end{equation}
$\alpha_{\mathbf{q}}$ is given by
\begin{equation}
\alpha_{\mathbf{q}} = -\gamma_{\mathbf{q}} \varphi_{\mathbf{q}} - \delta_{\mathbf{q}} \varphi_{\mathbf{-q}}^*,
\label{alpha}
\end{equation}
where we have introduced the following symbols,
\begin{eqnarray}
\gamma_{\mathbf{q}} = \frac{\omega_{\mathbf{-q}}}{\omega_{\mathbf{q}} \omega_{\mathbf{-q}} - \Delta^2}, \:\:\: \delta_{\mathbf{q}} = - \frac{\Delta}{\omega_{\mathbf{q}} \omega_{\mathbf{-q}} - \Delta^2}.
\label{gamma_def}
\end{eqnarray}
$\alpha_{\mathbf{q}}$ is determined self-consistently from Eqs. (\ref{phi_def}) and (\ref{alpha}) as a function of $V_{\mathbf{q}}$.

The momentum distributions are calculated as
\begin{eqnarray}
\rho(\mathbf{q}) = u^2_{\mathbf{q}} \pi_{\mathbf{q}}+v^2_{\mathbf{q}} (\pi_{\mathbf{-q}}+1) + \Left< |\alpha_{\mathbf{q}}|^2 \Right>,
\label{rho}
\end{eqnarray}
\begin{eqnarray}
\sigma(\mathbf{q}) = u_{\mathbf{q}} v_{\mathbf{q}} (\pi_{\mathbf{q}} + \pi_{\mathbf{-q}} +1) + \Left< \alpha_{\mathbf{q}} \alpha_{\mathbf{-q}} \Right>,
\label{sigma}
\end{eqnarray}
where $ \pi_{\mathbf{q}} = \langle \hat{c}^{\dag}_{\mathbf{q}} \hat{c}_{\mathbf{q}} \rangle = \{ \exp(E_{\mathbf{q}}/T)-1 \}^{-1}$.
We define
\begin{equation}
\rho_{\rm N}(\mathbf{q}) = u^2_{\mathbf{q}} \pi_{\mathbf{q}}+v^2_{\mathbf{q}} (\pi_{\mathbf{-q}}+1),
\end{equation}
\begin{equation}
\rho_{\rm G}(\mathbf{q}) = \Left< |\alpha_{\mathbf{q}}|^2 \Right>,
\label{rho_G}
\end{equation}
\begin{equation}
\sigma_{\rm N}(\mathbf{q}) = u_{\mathbf{q}} v_{\mathbf{q}} (\pi_{\mathbf{q}} + \pi_{\mathbf{-q}} +1),
\end{equation}
\begin{equation}
\sigma_{\rm G}(\mathbf{q}) = \Left< \alpha_{\mathbf{q}} \alpha_{\mathbf{-q}} \Right>.
\label{sigma_G}
\end{equation}
We call $\rho_{\rm G}(\mathbf{q})$ and $\sigma_{\rm G}(\mathbf{q})$ the glassy components, since they are analogous to the Edwards-Anderson parameter in spin glass theory \cite{Yukalov07-1}.
We also define the normal fraction and glassy fraction as
\begin{equation}
n_{\rm N,G} = \frac{1}{N_s} \sum_{\mathbf{q} \neq \mathbf{0}} \rho_{\rm N,G}(\mathbf{q}), \:\:\: \sigma_{\rm N,G} = \frac{1}{N_s} \sum_{\mathbf{q} \neq \mathbf{0}} \sigma_{\rm N,G}(\mathbf{q}),
\label{n1_sigma1}
\end{equation}
respectively. Note that $n_1=n_{\rm N}+n_{\rm G}$ and $\sigma_1=\sigma_{\rm N}+\sigma_{\rm G}$.

Eqs.~(\ref{Delta_def}), (\ref{phi_def}), (\ref{alpha}), (\ref{gamma_def}), (\ref{rho}), and (\ref{sigma}) compose a set of self-consistent equations.
This set can be solved as follows:
For a fixed $ \Delta $, $ \gamma_{\mathbf{q}} $ and $ \delta_{\mathbf{q}} $ are calculated using Eq.~(\ref{gamma_def}).
By solving Eqs.~(\ref{phi_def}) and (\ref{alpha}), we obtain $ \alpha_{\mathbf{q}} $ as a functional of $ V_{\mathbf{q}} $.
Then, we have $n_0=n-n_1$ and $\sigma_1$ as functions of $ \Delta $ from Eqs.~(\ref{rho}) and (\ref{sigma}).
Thus, Eq.~(\ref{Delta_def}) becomes a nonlinear equation with respect to $ \Delta $.
By solving this equation, we can calculate $ \Delta $, $n_0$, and $\rho(\mathbf{q})$ as functions of the strength of the disorder $V_0$ and the flow velocity $K$.

Next, let us calculate the chemical potential of the condensed particles $\mu_0$, the current density $j$, and the superfluid density $n_{\mathrm{s}}$ for later convenience.
From Eq.~(\ref{mu_0_def}), we have
\begin{equation}
\mu_0 = \epsilon_{K} + (n_0+2n_1+\sigma_1)U + \frac{1}{2 \sqrt{n_0} N_s} \sum_{\mathbf{q} \neq \mathbf{0}} \Left< V_{\mathbf{q}} \alpha^{*}_{\mathbf{q}} + V_{\mathbf{q}}^* \alpha_{\mathbf{q}} \Right>.
\end{equation}
By using Eqs.~(\ref{phi_def}) and (\ref{alpha}), the last terms are calculated as $\Left< V_{\mathbf{-q}} \alpha_{\mathbf{q}} \Right> \simeq -\sqrt{n_0}(\gamma_{\mathbf{q}}+\delta_{\mathbf{q}}) V_0^2$.
Thus, $\mu_0$ is given by
\begin{equation}
\mu_0 = \epsilon_{K} + (n+n_1+\sigma_1)U - \frac{1}{N_s} \sum_{\mathbf{q} \neq \mathbf{0}} (\gamma_{\mathbf{q}}+\delta_{\mathbf{q}}) V_0^2.
\label{mu_0}
\end{equation}
The current density operator is defined by
\begin{equation}
\hat{j} = i \frac{J}{N_s} \sum_i (\hat{b}^{\dag}_{\mathbf{R_i+e_x}} \hat{b}_{\mathbf{R_i}} - \hat{b}^{\dag}_{\mathbf{R_i}} \hat{b}_{\mathbf{R_i+e_x}}).
\end{equation}
From Eq.~(\ref{Bogoliubov_shift}), the averaged current density $j$ is given by
\begin{eqnarray}
j &=& \langle \hat{j} \rangle = 2J n_0 {\rm sin}K + \frac{1}{N_s} \sum_{\mathbf{q} \neq \mathbf{0}} 2J{\rm sin}(K+q_x) \rho(\mathbf{q}) \nonumber \\
&=& 2J n {\rm sin}K + \frac{1}{N_s} \sum_{\mathbf{q} \neq \mathbf{0}} 2J \bigl\{ {\rm sin}(K+q_x)-{\rm sin}K \bigr\} \rho(\mathbf{q}).
\label{j_def}
\end{eqnarray}
We define the superfluid density $n_{\rm s}$ in the presence of a finite flow by using the averaged current density Eq.~(\ref{j_def}):
\begin{eqnarray}
n_{\rm s} = \frac{j}{2J {\rm sin}K} = n + \frac{N_s^{-1} \sum_{\mathbf{q} \neq \mathbf{0}} \{{\rm sin}(K+q_x)-{\rm sin} K \} \rho(\mathbf{q}) }{ {\rm sin} K },
\label{ns_def}
\end{eqnarray}
which is simply the response of the current density with respect to a velocity boost.
The conventional superfluid density at equilibrium can be obtained by taking the limit $K \to 0$ in Eq.~(\ref{ns_def}).

\subsection{Landau and dynamical instability}

In order to derive the conditions for Landau and dynamical instability, we use the hydrodynamic approach \cite{Machholm}.
It is valid for an excitation for which the wavelength is much larger than the correlation length of the random potential.
Let us introduce an average particle density $ \bar{n}(\bf{r}) $, an average phase $ \bar{\phi}(\bf{r}) $, and an average momentum $ \bar{\bf{K}}(\bf{r}) = \nabla \bar{\phi}(\bf{r}) $, where the average is taken over a volume having linear dimensions much larger than the correlation length of the random potential.
We now derive the hydrodynamic equations for $ \bar{n}(\bf{r}) $ and $ \bar{K}(\bf{r}) $.
Let $ \mu_0(\bar{n},\bar{\bf{K}}) $ and $ \mathbf{j} (\bar{n},\bar{\bf{K}}) $ be the chemical potential of the condensed particles and the current density, respectively.
The phase evolves according to the Josephson equation, 
$\partial \bar{\phi}/\partial t = -\mu_0$,
which leads to the equation for the average momentum,
\begin{equation}
\frac{\partial \bar{\bf{K}}}{\partial t} = -\nabla \mu_0(\bar{n},\bar{\bf{K}}).
\label{hydro_eq_K}
\end{equation}
We also have the equation of continuity,
\begin{equation}
\frac{\partial \bar{n}}{\partial t} = -\nabla \cdot \mathbf{j}(\bar{n},\bar{\bf{K}}).
\label{hydro_eq_n}
\end{equation}

In order to obtain the hydrodynamic excitations, we linearize Eqs. (\ref{hydro_eq_K}) and (\ref{hydro_eq_n}).
For simplicity, we will consider the one-dimensional case.
For a steady state $ \bar{n}(x) = n$ and $ \bar{K}(x) = K$, we denote small changes in the density by $ \delta \bar{n}(x) $ and those in the momentum by $ \delta \bar{K}(x) $.
If one considers solutions varying in space and time as ${\rm exp} [i(qx-\Omega_q t)]$, one finds that 
\begin{equation}
\left( \frac{\partial \mu_0}{\partial K} q-\Omega_q \right) \delta \bar{K} + \frac{\partial \mu_0}{\partial n} q \delta \bar{n} = 0,
\end{equation}
\begin{equation}
\frac{\partial j}{\partial K} q \delta \bar{K} + \left( \frac{\partial j}{\partial n} q - \Omega_q \right) \delta \bar{n} = 0,
\end{equation}
where the derivatives are to be evaluated for the steady state.
From the above equations, we obtain the spectrum of hydrodynamic excitation,
\begin{eqnarray}
\Omega_{\mathbf{q}} = \frac{1}{2} \left( \frac{\partial \mu_0}{\partial K} + \frac{\partial j}{\partial n} \right) q_x + \frac{1}{2} \left\{ \left( \frac{\partial \mu_0}{\partial K}-\frac{\partial j}{\partial n} \right)^2 + 4 \frac{\partial \mu_0}{\partial n} \frac{\partial j}{\partial K} \right\}^{1/2} |\mathbf{q}|.
\label{spec_collec}
\end{eqnarray}
When $V_0=0$ and $T=0$, we have $ \mu_0 = \partial e(n,K)/\partial n $ and $ j = \partial e(n,K)/\partial K $, 
where $ e(n,k) \equiv \langle H \rangle/N_s $ is the total energy density.
Then, we obtain
\begin{eqnarray}
\Omega_{\mathbf{q}} = \frac{\partial^2 e}{\partial K \partial n}  q_x + \left( \frac{\partial^2 e}{\partial n^2} \frac{\partial^2 e}{\partial K^2} \right)^{1/2} |\mathbf{q}|,
\label{spec_collec_uniform}
\end{eqnarray}
which is also derived in Ref.~\cite{Machholm}.
Although Eq.~(\ref{spec_collec_uniform}) is useful for the weak disorder and low temperature regime, in what follows, we use the general formula Eq.~(\ref{spec_collec}).
Note that $\Omega_{\mathbf{q}}$, which is the spectrum of hydrodynamic collective excitation, differs from $E_q$ in Eq.~(\ref{E_q}), which is the spectrum of one-particle excitation.

When $K=0$, $\Omega_{\mathbf{q}}$ is given by
\begin{eqnarray}
\Omega_{\mathbf{q}} = \left( \frac{\partial \mu_0}{\partial n} \frac{\partial j}{\partial K} \right)^{1/2} |\mathbf{q}|,
\end{eqnarray}
because $\partial \mu_0/\partial K, \: \partial j/\partial n \propto K$ for small $K$.
Thus, the sound velocity at equilibrium is given by 
\begin{equation}
s_0 = \left( \frac{\partial \mu_0}{\partial n} \frac{\partial j}{\partial K} \right)^{1/2},
\end{equation}
where the derivatives are to be evaluated at $K=0$.
In the absence of disorder, the sound velocity $s_0$ should be identical to $c_0=\sqrt{ 2J \Delta}$, which is obtained from the spectrum of one-particle excitation $E_{\mathbf{q}}$ in Eq.~(\ref{E_q}), at zero temperature \cite{Gavoret}.
However, note that within the framework of our theory the sound velocity corresponding to hydrodynamic excitation does not exactly agree with that corresponding to one-particle excitation even if $V_0=0$ and $T=0$.
Although this is an artifact of the HFB approximation, the deviation between them is quite small for the weak interaction.

Landau instability occurs when there exists a wave number $\mathbf{q}$ such that the spectrum $\Omega_{\mathbf{q}}$ is negative.
Thus, the condition for the onset of Landau instability is written as
\begin{eqnarray}
\frac{\partial \mu_0}{\partial K} \frac{\partial j}{\partial n} = \frac{\partial \mu_0}{\partial n} \frac{\partial j}{\partial K}.
\label{Landau_condition}
\end{eqnarray}
Dynamical instability occurs when there exists a wave number $\mathbf{q}$ such that the spectrum $\Omega_{\mathbf{q}}$ has an imaginary part.
Thus, the condition for the onset of dynamical instability is written as
\begin{eqnarray}
\left( \frac{\partial \mu_0}{\partial K}-\frac{\partial j}{\partial n} \right)^2 + 4 \frac{\partial \mu_0}{\partial n} \frac{\partial j}{\partial K} =0.
\label{Dynamical_condition}
\end{eqnarray}
By solving the set of self-consistent equations given in Sec.~\ref{sec:Self-consistent equations} and using Eqs.~(\ref{mu_0}) and (\ref{j_def}), we can obtain $ \mu_0(n,K) $ and $ j(n,K) $.
From the conditions given in Eqs.~(\ref{Landau_condition}) and (\ref{Dynamical_condition}), we can calculate the critical velocities as functions of the strength of the random potential $V_0$ and temperature $T$.

\section{Results}
\label{sec:Results}

\subsection{Approximation}

In this section, we numerically solve the set of self-consistent equations derived in Sec.~\ref{sec:Self-consistent equations}.
Since it is hard to solve Eqs.~(\ref{phi_def}) and (\ref{alpha}), we employ some approximations.
First, the effective random potential $\varphi_{\mathbf{q}}$ is approximated to the second order:
\begin{equation}
\varphi_{\mathbf{q}} = \sqrt{n_0} V_{\mathbf{q}} - \frac{1}{\sqrt{N_s}} \sum_{\mathbf{q'} \neq \mathbf{0}} \sqrt{n_0} (\gamma_{\mathbf{q'}} + \delta_{\mathbf{q'}}) V_{\mathbf{q'}} V_{\mathbf{q-q'}}. 
\end{equation}
Then, we have
\begin{eqnarray}
\Left< |\varphi_{\mathbf{q}}|^2 \Right> = n_0 V_0^2 + \frac{1}{N_s} \sum_{\mathbf{q'} \neq \mathbf{0}} n_0 V_0^4 \bigl\{ (\gamma_{\mathbf{q'}} + \delta_{\mathbf{q'}})^2 + (\gamma_{\mathbf{q'}} + \delta_{\mathbf{q'}})(\gamma_{\mathbf{q-q'}} + \delta_{\mathbf{q-q'}}) \bigr\},
\end{eqnarray}
\begin{eqnarray}
\Left< \varphi_{\mathbf{q}} \varphi_{\mathbf{-q}} \Right> = n_0 V_0^2 + \frac{1}{N_s} \sum_{\mathbf{q'} \neq \mathbf{0}} n_0 V_0^4 \bigl\{ (\gamma_{\mathbf{q'}} + \delta_{\mathbf{q'}})(\gamma_{\mathbf{-q'}} + \delta_{\mathbf{-q'}}) + (\gamma_{\mathbf{q'}} + \delta_{\mathbf{q'}})(\gamma_{\mathbf{q'-q}} + \delta_{\mathbf{q'-q}}) \bigr\}.
\end{eqnarray}
Furthermore, we ignore the $q$ dependence of $ \Left< |\varphi_{\mathbf{q}}|^2 \Right> $ and $ \Left< \varphi_{\mathbf{q}} \varphi_{\mathbf{-q}} \Right> $, and they are replaced with $ \Left< |\varphi_{0}|^2 \Right> \equiv \lim_{\mathbf{q} \to \mathbf{0}} \Left< |\varphi_{\mathbf{q}}|^2 \Right> $ and $ \Left< \varphi_{0}^2 \Right> \equiv \lim_{\mathbf{q} \to \mathbf{0}} \Left< \varphi_{\mathbf{q}} \varphi_{\mathbf{-q}} \Right> $, respectively.
This simplification is justified by assuming that in the summation in Eq.~(\ref{n1_sigma1}), the main contributions come from the region of small momenta.
Thus, we have
\begin{eqnarray}
\Left< |\varphi_{0}|^2 \Right> = \Left< \varphi_{0}^2 \Right> = n_0 V_0^2 + \frac{1}{N_s} \sum_{\mathbf{q'} \neq \mathbf{0}} n_0 V_0^4 \bigl\{ (\gamma_{\mathbf{q'}} + \delta_{\mathbf{q'}})^2 + (\gamma_{\mathbf{q'}} + \delta_{\mathbf{q'}})(\gamma_{\mathbf{-q'}} + \delta_{\mathbf{-q'}}) \bigr\}.
\end{eqnarray}
By using this approximation, $\rho_{\rm G}(\mathbf{q})$ and $\sigma_{\rm G}(\mathbf{q})$ are calculated as
\begin{eqnarray}
\rho_{\rm G}(\mathbf{q})&=&(\gamma_{\mathbf{q}}+\delta_{\mathbf{q}})^2 \Left< |\varphi_{0}|^2 \Right>, \nonumber \\
\sigma_{\rm G}(\mathbf{q})&=&(\gamma_{\mathbf{q}}+\delta_{\mathbf{q}}) (\gamma_{\mathbf{-q}}+\delta_{\mathbf{-q}}) \Left< \varphi_{0}^2 \Right>.
\end{eqnarray}
If we introduce simple notations
\begin{eqnarray}
A(\Delta) &=& \frac{1}{N_s} \sum_{\mathbf{q} \neq \mathbf{0}} (\gamma_{\mathbf{q}}+\delta_{\mathbf{q}})^2, \nonumber \\
B(\Delta) &=& \frac{1}{N_s} \sum_{\mathbf{q} \neq \mathbf{0}} (\gamma_{\mathbf{q}}+\delta_{\mathbf{q}})(\gamma_{\mathbf{-q}}+\delta_{\mathbf{-q}}),
\end{eqnarray}
we have $n_0$ as a function of $\Delta$,
\begin{eqnarray}
n_0(\Delta) = \bigl( n-n_{\rm N}(\Delta) \bigr) \bigl[ 1 + A(\Delta)V_0^2 + A(\Delta) \bigl\{ A(\Delta)+B(\Delta) \bigr\} V_0^4 \bigr]^{-1},
\label{n0-approx}
\end{eqnarray}
where we have used $n_{\rm G} = N_s^{-1} \sum_{\mathbf{q} \neq \mathbf{0}} \rho_{\rm G}(\mathbf{q})$ and $n_0+n_{\rm N}+n_{\rm G}=n$.
By using this equation, $\sigma_1$ is also calculated as
\begin{eqnarray}
\sigma_1(\Delta) = \sigma_{\rm N}(\Delta) + n_0(\Delta) V_0^2 B(\Delta) \bigl[ 1 + \bigl\{ A(\Delta)+B(\Delta) \bigr\} V_0^2 \bigr].
\label{sigma-approx}
\end{eqnarray}
Insertion of Eqs.~(\ref{n0-approx}) and (\ref{sigma-approx}) into Eq.~(\ref{Delta_def}) leads to a self-consistent equation of $\Delta$.
Once $\Delta$ is determined, the other quantities can be also calculated.

In the following calculations, we consider a two-dimensional system with an interaction $U=0.2$ and total density $n=1$.
Interaction energy $U$, strength of the random potential $V_0$, and temperature $T$ are measured in units of $J$.

\subsection{Equilibrium case}

%%%%%%%%%%%%%%%%%%%%%%%%%%%%%%%%%%%%%%%%%%%%%%%%%%%%%%%
\begin{figure}
  \centering
  \includegraphics[width=0.6\textwidth]{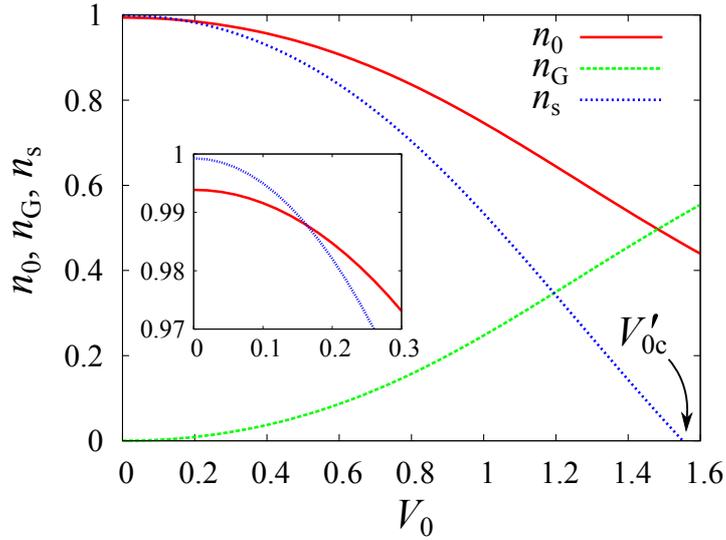}
\caption{$n_0$, $n_{\rm G}$, and $n_{\rm s}$ as functions of $V_0$, when $K=0$. 
 The system size is $L=50$ and the temperature is $T=0$.
 The inset displays the small-$V_0$ region of the same graph (Color figure online)}
 \label{fig:n-V0-flowless}
\end{figure}
%%%%%%%%%%%%%%%%%%%%%%%%%%%%%%%%%%%%%%%%%%%%%%%%%%%%%%%

First, we consider the case in which $K=0$.
Fig.~\ref{fig:n-V0-flowless} shows the condensate density $n_0$, the glassy fraction $n_{\rm G}$, and the superfluid density $n_{\rm s}$ as functions of the strength of the random potential $V_0$ at zero temperature.
The system size is $L=50$.
$n_{\rm s}$ is calculated by taking the limit $K \to 0$ in Eq.~(\ref{ns_def}).
$n_0$ and $n_{\rm s}$ decrease with increasing $V_0$; eventually, $n_{\rm s}$ vanishes at $V_0=V_{\rm 0c}'$.
Although this behavior is reminiscent of the superfluid--Bose-glass transition, we emphasize that the Bogoliubov theory is incapable of describing this transition.
Since the superfluid--Bose-glass transition is caused by the percolation of localized condensates, it cannot be captured in the framework of the Bogoliubov theory, which is a perturbation theory around a single macroscopic condensate.
In fact, in the strong disorder regime, the behaviors shown in Fig.~\ref{fig:n-V0-flowless} are inconsistent with those found in the previous studies. 
At the superfluid--Bose-glass transition point, $n_0$ and $n_{\rm s}$ should vanish simultaneously \cite{Zuniga}, while in our calculation, $n_0$ has a finite value at $V_{\rm 0c}'$.
Furthermore, $V_{\rm 0c}' \simeq 1.55$ is much lower than the actual transition point $V_{\rm 0c} \simeq 20\sqrt{Ut} \simeq 9 $, which was obtained from a quantum Monte Carlo (QMC) study \cite{Soyler}.
In this QMC study, it was observed that in the strong disorder regime, there is a wide region in which $n_0/n$ and $n_{\rm s}/n$ have extremely small values.
The Bogoliubov theory breaks down in such a fragile superfluid state, which is why $V_{\rm 0c}'$ is much lower than $V_{\rm 0c}$.
However, we can expect that our theory is correct for weak and moderate disorder.
In the small-$V_0$  regime, $n_{\rm s}$ is larger than $n_0$, while in the large-$V_0$ regime, $n_{\rm s}$ falls below $n_0$.
This behavior was pointed out in early studies \cite{Huang}.
In our calculation, $n_{\rm s}$ is slightly smaller than $n=1$ at $V_0=0$ because the dispersion relation $\epsilon_q$ is not completely parabolic.

\subsection{Dynamical phase diagram}

%%%%%%%%%%%%%%%%%%%%%%%%%%%%%%%%%%%%%%%%
\begin{figure}
 \centering
 \includegraphics[width=0.6\textwidth]{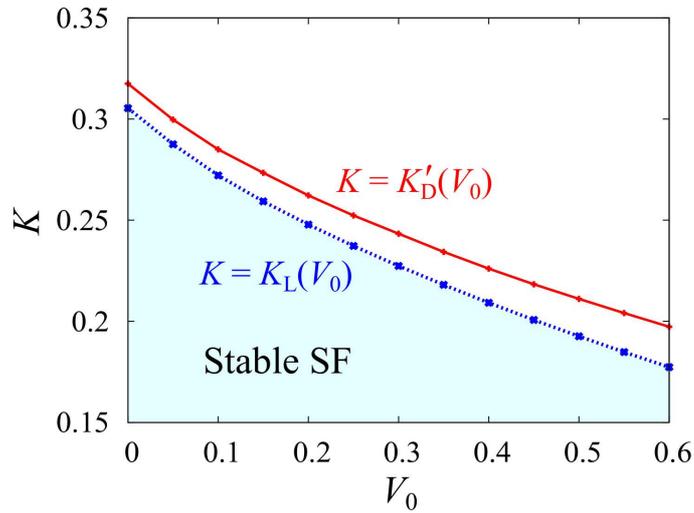}
 \caption{Dynamical phase diagram with respect to $V_0$ and $K$.
 The system size is $L=30$ and the temperature is $T=0$.
 At the solid line (red), the solution of the self-consistent equation disappears.
 The dotted line (blue) represents the onset of Landau instability (Color figure online)}
 \label{fig:phase-diagram}
\end{figure}
%%%%%%%%%%%%%%%%%%%%%%%%%%%%%%%%%%%%%%%%

Next, we consider the cases in which $K \neq 0$.
One can distinguish three ``critical'' momenta, which are denoted as $K_{\rm L}$, $K_{\rm D}$, and $K_{\rm D}'$.
$K_{\rm L}$ is the critical momentum corresponding to Landau instability, which is defined by Eq.~(\ref{Landau_condition}).
$K_{\rm D}$ is the critical momentum corresponding to dynamical instability, which is defined by Eq.~(\ref{Dynamical_condition}).
$K_{\rm D}'$ is the critical momentum above which the self-consistent equation has no solution.
We expect that these critical momenta satisfy $K_{\rm L} < K_{\rm D} < K_{\rm D}'$.
Fig.~\ref{fig:phase-diagram} shows the dynamical phase diagram with respect to the strength of the random potential $V_0$ and the flow momentum $K$ at zero temperature.
The system size is $L=30$.
The red solid line and the blue dotted line represent $K=K_{\rm D}'(V_0)$ and $K=K_{\rm L}(V_0)$, respectively.
It is worth noting that $K_{\rm D}'(V_0=0) \neq \lim_{V_0 \to 0} K_{\rm D}'(V_0)$ and $K_{\rm D}(V_0=0) \neq \lim_{V_0 \to 0} K_{\rm D}(V_0)$.
In fact, if we consider a continuous system in which the kinetic energy of a free particle is given by a parabolic function, the self-consistent solution exists for an arbitrary velocity in the absence of the disorder; $K_{\rm D}'(V_0=0)=\infty$, while $\lim_{V_0 \to 0} K_{\rm D}'(V_0)$ is finite.
This is also true for $K_{\rm D}(V_0)$.

We have confirmed that the following relation holds,
\begin{equation}
\frac{\partial \mu_0}{\partial K} = \frac{\partial j}{\partial n},
\end{equation}
with high accuracy.
Thus, the condition for the onset of dynamical instability Eq.~(\ref{Dynamical_condition}) can be reduced to
\begin{equation}
\frac{\partial j}{\partial K} =0,
\end{equation}
which means that the inverse of the effective mass vanishes \cite{Machholm}.
We found that $K_{\rm D}$ is very close to $K_{\rm D}'$.
Thus, in the following, we consider $K_{\rm D}'$ instead of $K_{\rm D}$.
For sufficiently low temperatures, since dynamical instability plays a dominant role in the breakdown of superfluidity \cite{Sarlo,Iigaya}, $K=K_{\rm D}'(V_0)$ approximates the observable critical velocity.
It is worth noting that $K_{\rm D}'(V_0)$ decreases linearly in the small-$V_0$ regime, $K_{\rm D}'(V_0)=K_{\rm D}'(0)-c V_0$.
This behavior contrasts with that of superfluid Bose gases flowing past an obstacle in one dimension, in which the critical momentum drops to $K_{\rm c}=K_{\rm c0}-c g^{2/3}$, where $g$ is the strength of the obstacle potential \cite{Hakim}.

$K_{\rm L}$ and $K_{\rm D}'$ are expected to vanish at the superfluid--Bose-glass transition point $V_0=V_{0\mathrm{c}}$.
In fact, it was demonstrated in numerical simulations that, near this transition point, the superfluid flow breaks down even when the current is quite small \cite{Buonsante15}.

%%%%%%%%%%%%%%%%%%%%%%%%%%%%%%%%%%%%%%%%
\begin{figure}
 \centering
 \includegraphics[width=0.9\textwidth]{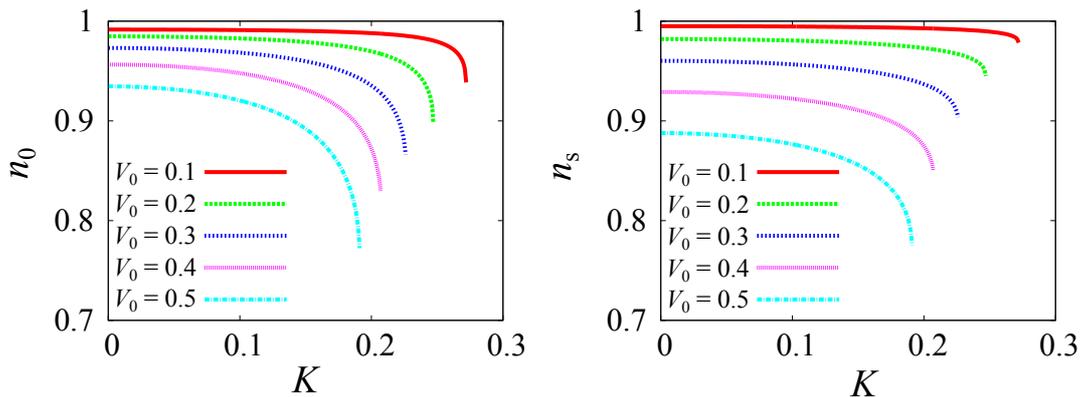}
 \caption{Condensate density $n_0$ and superfluid density $n_{\rm s}$ 
 as functions of $K$ for different disorder strengths.
 The values of $V_0$ are $V_0=0.1$, $0.2$, $0.3$, $0.4$ and $0.5$.
 $n_0$ and $n_{\rm s}$ are plotted up to $K_{\rm D}'(V_0)$.
 The system size is $L=50$ and the temperature is $T=0$ (Color figure online)}
 \label{fig:n-K}
\end{figure}
%%%%%%%%%%%%%%%%%%%%%%%%%%%%%%%%%%%%%%%%

Fig.~\ref{fig:n-K} shows the condensate density $n_0$ and the superfluid density $n_{\rm s}$ as functions of $K$ for weak and moderate disorder; $n_0$ and $n_{\rm s}$ are plotted up to $K_{\rm D}'(V_0)$.
They decrease with increasing $K$ because the condensed particles are more strongly scattered by the random potential.
Note that $n_0$ and $n_{\rm s}$ have non-zero values at $K=K_{\rm D}'(V_0)$.
Above the critical momentum, the steady flow becomes unstable, and we speculate that this is where the transition to turbulent flow occurs.
This transition is not thermodynamic, and so $n_0$ and $n_{\rm s}$ do not necessarily vanish, in contrast to the second-order transition at equilibrium.

\subsection{System size dependence of the critical momenta}

\begin{figure}
 \centering
 \includegraphics[width=0.9\textwidth]{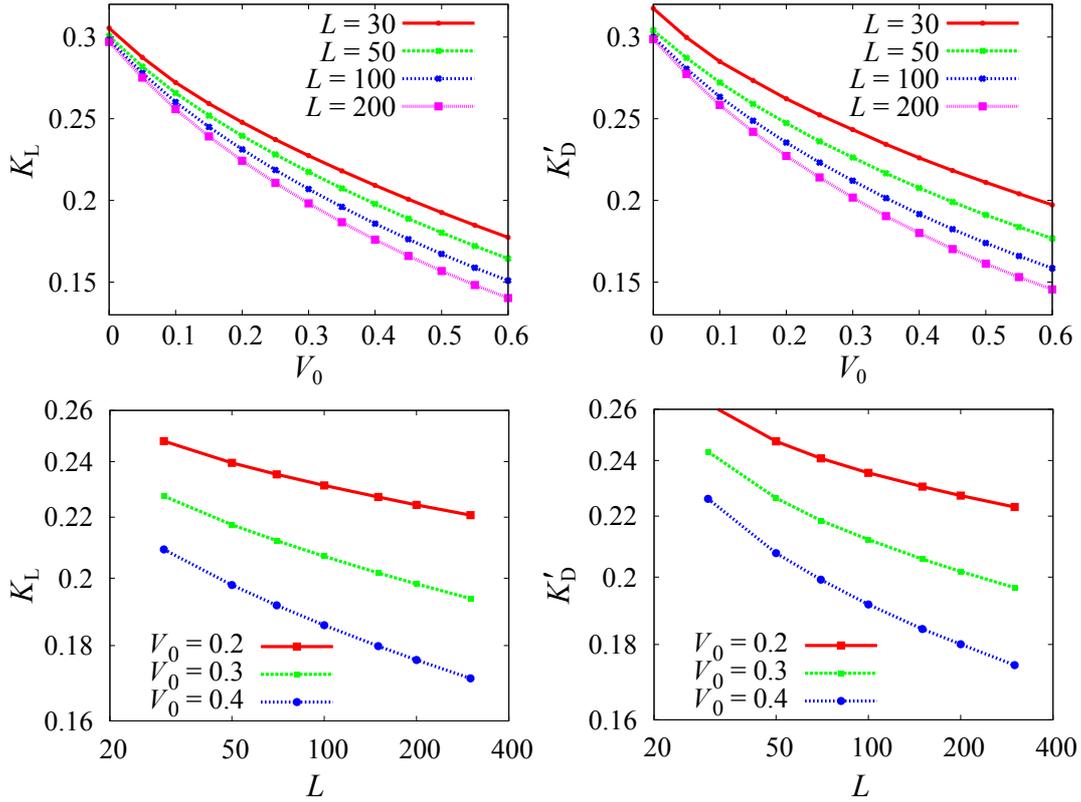}
 \caption{(Upper panels): $K_{\rm L}$ and $K_{\rm D}'$ as functions of $V_0$ for different system sizes.
 The system sizes are $L=30$, $50$, $100$, and $200$, and the temperature is $T=0$. 
 (Lower panels): $K_{\rm L}$ and $K_{\rm D}'$ as functions of $L$ for different values of the disorder strength.
 They are plotted in log scale on the both axes.
 The values of the disorder strength are $V_0=0.2$, $0.3$ and $0.4$, and the temperature is $T=0$ (Color figure online)}
 \label{fig:K-size}
\end{figure}

We investigate the system size dependence of the critical momenta $K_{\rm L}$ and $K_{\rm D}'$ for a fixed particle density.
The upper panels in Fig.~\ref{fig:K-size} show $K_{\rm L}$ and $K_{\rm D}'$ for different system sizes at zero-temperature.
Note that the values of $\lim_{V_0 \to 0} K_{\rm L}(V_0)$ and $\lim_{V_0 \to 0} K_{\rm D}'(V_0)$ converge to finite limits as $L$ goes to infinite.
When $V_0>0$, $K_{\rm L}$ and $K_{\rm D}'$ decrease as the system size $L$ increases.
One can see that the critical momenta exhibit strong system size dependence even for large system sizes.
Furthermore, $K_{\rm L}(V_0)$ and $K_{\rm D}'(V_0)$ become closer and closer with increasing $L$.
In fact, for $L=200$, the deviation $|K_{\rm D}'(V_0)-K_{\rm L}(V_0)|/K_{\rm D}'(V_0)$ is the order of $0.01$.

%%%%%%%%%%%%%%%%%%%%%%%%%%%%%%%%%%%%%%%%
%%%%%%%%%%%%%%%%%%%%%%%%%%%%%%%%%%%%%%%%

We now consider whether the critical momenta $K_{\rm L}(V_0)$ and $K_{\rm D}'(V_0)$ converge to finite values in the thermodynamic limit $L \to \infty$.
The lower panels in Fig.~\ref{fig:K-size} show $K_{\rm L}$ and $K_{\rm D}'$ as functions of the system size for different values of the disorder strength.
These graphs suggest a logarithmic dependence with respect to the system size,
\begin{equation}
K_{\mathrm{c}}(V_0,L) \simeq \tilde{K}_{\mathrm{c}}(V_0) \bigl( 1-f(V_0) \ln L \bigr) \simeq \tilde{K}_{\mathrm{c}}(V_0) L^{-f(V_0)},
\label{K_log_dependence}
\end{equation}
where $f(V_0)$ is an increasing function satisfying $f(0)=0$.
This logarithmic dependence can be understood as follows.
By using Eqs.~(\ref{phi_def}), (\ref{alpha}), and (\ref{rho_G}), we  obtain
\begin{equation}
\rho_{\rm G}(\mathbf{q})=(\gamma_{\mathbf{q}}+\delta_{\mathbf{q}})^2 n_0 V_0^2 + O(V_0^4),
\label{rho_G-second-order}
\end{equation}
for weak disorder.
When $K=0$, from Eq.~(\ref{gamma_def}), we obtain
\begin{equation}
\rho_{\rm G}(\mathbf{q})=\frac{1}{(\omega_{\mathbf{q}} + \Delta)^2} n_0 V_0^2 + O(V_0^4),
\end{equation}
which is regular at $\mathbf{q} = \mathbf{0}$.
Thus, $n_{\rm G} = N_s^{-1} \sum_{\mathbf{q} \neq \mathbf{0}} \rho_{\rm G}(\mathbf{q})$ converges to a finite value in the thermodynamic limit $L \to \infty$.
In contrast, when $K \neq 0$, $\gamma_{\mathbf{q}}$ and $\delta_{\mathbf{q}}$ behave near $\mathbf{q} = \mathbf{0}$ as
\begin{eqnarray}
\gamma_{\mathbf{q}} \simeq \frac{1}{2J \Delta} \frac{-2KJq_x + \Delta}{(1-K^2/K_{\mathrm s}^2)q_x^2 + q_{\perp}^2} + \gamma_{\mathbf{q}}^{(0)}, \nonumber \\
\delta_{\mathbf{q}} \simeq \frac{1}{2J \Delta} \frac{-\Delta}{(1-K^2/K_{\mathrm s}^2)q_x^2 + q_{\perp}^2} + \delta_{\mathbf{q}}^{(0)},
\label{gamma_delta_asymp}
\end{eqnarray}
where $K_{\mathrm s}=\sqrt{\Delta/(2J)}$, $q_{\perp}^2 = q^2 - q_x^2$, and $\gamma_{\mathbf{q}}^{(0)}, \delta_{\mathbf{q}}^{(0)}$ are the nondivergent parts.
Note that $K_{\mathrm s}$ is the momentum corresponding to the sound velocity of the one-particle excitation Eq.~(\ref{E_q}).
Thus, from Eq.~(\ref{rho_G-second-order}), we find that $\rho_{\rm G}(\mathbf{q})$ has a singularity of the form $q^{-2}$:
\begin{equation}
\rho_{\rm G}(\mathbf{q}) = \rho_{\rm G}^{(0)}(\mathbf{q}) + \frac{(K/\Delta)^2 q_x^2}{\{ (1-K^2/K_{\mathrm s}^2)q_x^2 + q_{\perp}^2 \}^2} n_0 V_0^2 + O(V_0^4),
\label{rho_asymp}
\end{equation}
where $\rho_{\rm G}^{(0)}(\mathbf{q})$ is the nondivergent part of $O(V_0^2)$.
In two dimensions, this singularity $\rho_{\rm G}(\mathbf{q}) \sim q^{-2}$ leads to a logarithmic dependence,
\begin{equation}
n_{\rm G} = \frac{1}{N_s} \sum_{\mathbf{q} \neq \mathbf{0}}^{(2\mathrm{D})} \rho_{\rm G}(\mathbf{q}) \simeq n_{\rm G}^{(0)} + c(K^2) V_0^2 \ln L + O(V_0^4),
\end{equation}
where $c(K^2)$ is a function of $K^2$.
This logarithmic size dependence is carried over to the critical momenta, as shown in Eq.~(\ref{K_log_dependence}).
On the other hand, in three dimensions, since the integral of $\rho_{\rm G}(\mathbf{q})$ converges to a finite value in the limit $L \to \infty$, we expect that the critical momenta are independent of the system size.

If we accept Eq.~(\ref{K_log_dependence}) literally, the critical momenta should vanish in the thermodynamic limit, even when the disorder is infinitesimally weak,
\begin{eqnarray}
\lim_{V_0 \to 0} \lim_{L \to \infty} K_{\mathrm{c}}(V_0,L) &=& 0, \nonumber \\
\lim_{L \to \infty} \lim_{V_0 \to 0} K_{\mathrm{c}}(V_0,L) &=& \tilde{K}_{\mathrm{c}}(0) \neq 0.
\label{Kc_limit}
\end{eqnarray}
It is worth noting that the vanishing of the critical momenta does not contradict the fact that the system has a non-zero superfluid density at equilibrium.
For example, in one-dimensional Bose gas without disorder, the critical velocity is zero despite the presence of a non-zero superfluid density at zero-temperature \cite{Cherny}.

Within the framework of our theory, it is difficult to give a clear answer to the question whether the critical velocity of the two-dimensional disordered Bose gas vanishes in the thermodynamic limit.
Small critical velocity is a consequence of the low condensate fraction.
However, the Bogoliubov theory is invalid when the condensate fraction is extremely small, $n_0/n \ll 1$.
Thus, it is not reliable for large system sizes.
Furthermore, there is a possibility that the higher-order terms $O(V_0^4)$ in Eq.~(\ref{rho_asymp}) have stronger singularities of the form $q^{-2n}$ with $n \geq 2$.
The effect of these higher-order singularities, which are ignored in this calculation, is unclear at this time.
To establish the validity of Eqs.~(\ref{K_log_dependence}) and (\ref{Kc_limit}), further investigation is required.

\subsection{Finite temperature case}

\begin{figure}
 \centering
 \includegraphics[width=0.9\textwidth]{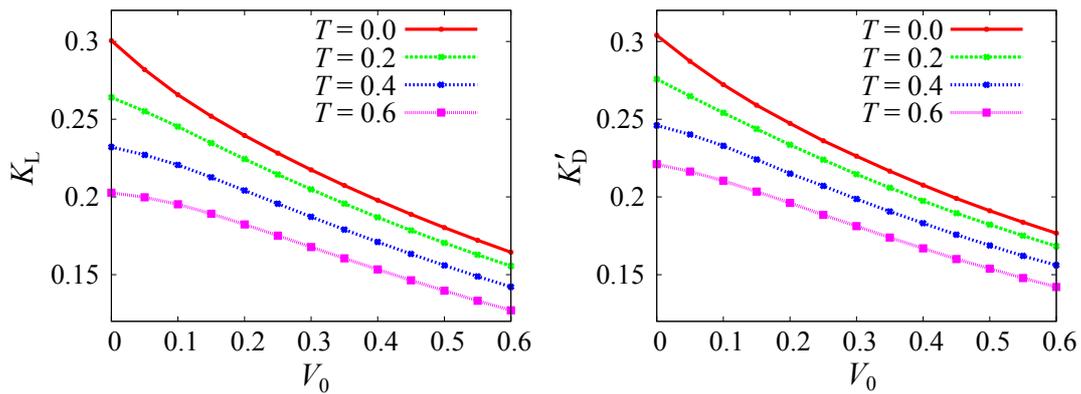}
 \caption{$K_{\rm L}$ and $K_{\rm D}'$ as functions of $V_0$ for different temperatures.
 The system size is $L=50$ and the temperatures are $T=0.0$, $0.2$, $0.4$, and $0.6$ (Color figure online)}
 \label{fig:K-temp}
\end{figure}

Finally, we consider the finite temperature case.
Fig.~\ref{fig:K-temp} shows the temperature dependence of $K_{\rm L}$ and $K_{\rm D}'$.
Since the thermal excitation reduces the condensate density, the critical velocity decreases with temperature.
It is worth noting that even in the absence of disorder, the Landau critical momentum $K_{\rm L}(V_0=0)$ is lower than the momentum corresponding to the sound velocity at equilibrium \cite{Navez}.
This fact can be understood as follows.
At zero-temperature, the spectrum of excitation for the moving state is given by Eq.~(\ref{Omega_move}) due to the Galilean invariance of the system.
At finite temperature, since the presence of a thermal cloud around the condensate breaks the Galilean invariance, the spectrum is no longer given by Eq.~(\ref{Omega_move}).
Thus, the Landau critical velocity can be different from the sound velocity at equilibrium.

%%%%%%%%%%%%%%%%%%%%%%%%%%%%%%%%%%%%%%%%
%%%%%%%%%%%%%%%%%%%%%%%%%%%%%%%%%%%%%%%%

\section{Summary and Discussions}
\label{sec:Summary}

This paper was devoted to the development of a general approach for treating a weakly interacting Bose gas flowing in a random potential.
By employing the Bogoliubov theory and the hydrodynamic approach, we discussed the stability of the steady flow in the presence of weak or moderate disorder.
The critical velocity, at which the steady flow becomes unstable, was calculated as a function of the disorder strength and temperature.
We found that the momentum distribution function of the noncondensed particles exhibits a singularity of the form $\rho(\mathbf{q}) \sim q^{-2}$ near $\mathbf{q} = 0$, even at zero-temperature.
This singular behavior leads to the strong system size dependence of the critical velocity in two dimensions.

In this study, the stable superfluid state is defined as the hydrodynamically stable state.
This means that this state is stable with respect to a hydrodynamic excitation whose wavelength is much larger than the correlation length of the random potential.
The hydrodynamically stable state is not necessarily dissipationless and time independent, because rugged local configurations of the random potential, whose length-scale is comparable to the correlation length of the random potential, can cause the nucleation of defects and lead to a finite amount of dissipation.

It is difficult to make a direct comparison of our results with the results of previous studies that dealt with one-dimensional Bose gases flowing in a random potential \cite{Albert,Paul}. 
In those studies, the critical velocity was defined as the flow velocity above which there exists no stable solution for the ``microscopic'' GP equation, which is the conventional GP equation with a single realization of the random potential $V(\mathbf{r})$.
Thus, one can distinguish two critical velocities: the microscopic critical velocity and the hydrodynamic critical velocity.
From the definition, the former is smaller than the latter.
The microscopic critical velocity is sensitive to each realization of $V(\mathbf{r})$ because it is closely related to an extreme value $\bar{V}=\max_{ \: \mathbf{r}} V(\mathbf{r})$.
Furthermore, for an unbounded disorder, it always vanishes in the thermodynamic limit because for larger system one has more chance to encounter rugged local configurations of the random potential, which cause the nucleation of defects. 
In contrast, for sufficiently large system sizes, the hydrodynamic critical velocity is almost independent of each realization of $V(\mathbf{r})$.
The critical velocity that can be observed in experiments is expected to be the hydrodynamic one.

It is known that the thermal activation process and quantum tunneling play crucial roles in the occurrence of phase slip, which is responsible for the onset of dissipation \cite{Polkovnikov,Danshita13}.
Because of such effects, the phase boundary of the stable superfluid can be smeared, and this broadening of the dynamical transition is substantial for one-dimensional systems.
Since the Bogoliubov theory just accounts for small fluctuations around the condensate, it cannot describe the phase slip process, in which the drastic change of the condensate wave function takes place.
Thus, our theory cannot predict the broadening of the dynamical transition resulting from the phase slip.

Finally, as mentioned in Sec.~\ref{sec:Bogoliubov_theory_with_disorder}, recall that there are two approaches in the Bogoliubov theory for a disordered Bose gas depending on how to decompose the field operator.
In the first scheme, the field operator is decomposed into a spatially deformed condensate and a quantum fluctuation, while in the second scheme, it is decomposed into a uniform background and a fluctuation around it.  
Not only thermodynamic quantities at equilibrium, but also the critical velocity can depend on the decomposition scheme.
For strong disorder, the first decomposition scheme is expected to be preferable.
In the future study, we will try to investigate the behavior of the critical velocity near the superfluid--Bose-glass transition point by employing this scheme.

\begin{acknowledgements}
The author thanks M. Kunimi and M. Kobayashi for fruitful discussions.
The present study was supported by Japan Society for the Promotion of Science (JSPS) KAKENHI Grant No. 15J01614, a Grant-in-Aid for JSPS Fellows.
\end{acknowledgements}


\begin{thebibliography}{}

%%% The breakdown of superfluidity: experiment liquid He %%%
\bibitem{Allum} D.R. Allum, P.V.E. McClintock, A. Phillips, Phil. Trans. R. Soc. A {\bf 284}, 179 (1977).
% The Breakdown of Superfluidity in Liquid $^{4}$He %

%%% The breakdown of superfluidity: experiment cold atom %%%
\bibitem{Chikkatur} A.P. Chikkatur, A. G\"{o}rlitz, D.M. Stamper-Kurn, S. Inouye, S. Gupta, and W. Ketterle, Phys. Rev. Lett. {\bf 85}, 483 (2000).
% Suppression and Enhancement of Impurity Scattering in a Bose-Einstein Condensate %
\bibitem{Inouye} S. Inouye, S. Gupta, T. Rosenband, A.P. Chikkatur, A. G\"{o}rlitz, T.L. Gustavson, A.E. Leanhardt, D.E. Pritchard, and W. Ketterle, Phys. Rev. Lett. {\bf 87}, 080402 (2001).
% Observation of Vortex Phase Singularities in Bose-Einstein Condensates %
\bibitem{Sarlo} L. De Sarlo, L. Fallani, J.E. Lye, M. Modugno, R. Saers, C. Fort, and M. Inguscio, Phys. Rev. A {\bf 72}, 013603 (2005).
% Unstable regimes for a Bose-Einstein condensate in an optical lattice %
\bibitem{Engels} P. Engels and C. Atherton, Phys. Rev. Lett. {\bf 99}, 160405 (2007).
% Stationary and Nonstationary Fluid Flow of a Bose-Einstein Condensate Through a Penetrable Barrier %
\bibitem{Mun} J. Mun, P. Medley, G.K. Campbell, L.G. Marcassa, D.E. Pritchard, and W. Ketterle, Phys. Rev. Lett. {\bf 99}, 150604 (2007).
% Phase Diagram for a Bose-Einstein Condensate Moving in an Optical Lattice %
\bibitem{Mckay} D. McKay, M. White, M. Pasienski, and B. DeMarco, Nature, {\bf 453}, 76 (2008).
% Phase-slip-induced dissipation in an atomic Bose-Hubbard system %
\bibitem{Neely} T.W. Neely, E.C. Samson, A.S. Bradley, M.J. Davis, and B.P. Anderson, Phys. Rev. Lett. {\bf 104}, 160401 (2010).
% Observation of Vortex Dipoles in an Oblate Bose-Einstein Condensate %

%%% The breakdown of superfluidity: theory (GP equation) %%%
\bibitem{Frisch} T. Frisch, Y. Pomeau, and S. Rica, Phys. Rev. Lett. {\bf 69}, 1644 (1992).
% Transition to dissipation in a model of superflow %
\bibitem{Hakim} V. Hakim, Phys. Rev. E {\bf 55}, 2835 (1997).
% Nonlinear Schrodinger flow past an obstacle in one dimension %
\bibitem{Winiecki} T. Winiecki, J.F. McCann, and C.S. Adams, Phys. Rev. Lett. {\bf 82}, 5186 (1999).
% Pressure Drag in Linear and Nonlinear Quantum Fluids %
\bibitem{Huepe} C. Huepe and M-E. Brachet, Physica D {\bf 140}, 126 (2000).
% Scaling laws for vortical nucleation solution in a model of superflow %
\bibitem{Pavloff} N. Pavloff, Phys. Rev. A {\bf 66}, 013610 (2002).
% Breakdown of superfluidity of an atom laser past an obstacle %
\bibitem{Danshita07} I. Danshita and S. Tsuchiya, Phys. Rev. A {\bf 75}, 033612 (2007).
% Stability of Bose-Einstein condensates in a Kronig-Penney potential %
\bibitem{Sasaki} K. Sasaki, N. Suzuki, and H. Saito, Phys. Rev. Lett. {\bf 104}, 150404 (2010).
% Benard-von Karman Vortex Street in a Bose-Einstein Condensate %
\bibitem{Kato} Y. Kato and S. Watabe, Phys. Rev. Lett. {\bf 105}, 035302 (2010).
% Dynamical density fluctuation of superfluids near the critical velocity %
\bibitem{Dubessy} R. Dubessy, T. Liennard, P. Pedri, and H. Perrin, Phys. Rev. A {\bf 86}, 011602(R) (2012).
% Critical rotation of an annular superfluid Bose-Einstein condensate %
\bibitem{Kunimi} M. Kunimi and Y. Kato, Phys. Rev. A {\bf 91}, 053608 (2015).
% Metastability, excitations, fluctuations, and multiple-swallowtail structures of a superfluid in a Bose-Einstein condensate in the presence of a uniformly moving defect %

%%% The Landau and dynamical stability %%%
\bibitem{Machholm} M. Machholm, C.J. Pethick, and H. Smith, Phys. Rev. A {\bf 67}, 053613 (2003).
% Band structure, elementary excitations, and stability of a Bose-Einstein condensate in a periodic potential %
\bibitem{Wu} B. Wu and Q. Niu, New J. Phys. {\bf 5}, 104 (2003).
% Superfluidity of Bose-Einstein condensate in an optical lattice %
\bibitem{Iigaya} K. Iigaya, S. Konabe, I. Danshita, and T. Nikuni, Phys. Rev. A {\bf 74}, 053611 (2006).
% Landau damping: Instability mechanism of superfluid Bose gases moving in optical lattices %

%%% Superfluid in a disordered environment:experiment %%%
\bibitem{Dries} D. Dries, S.E. Pollack, J.M. Hitchcock, and R.G. Hulet, Phys. Rev. A {\bf 82}, 033603 (2010).
% Dissipative transport of a Bose-Einstein condensate %
\bibitem{Tanzi} L. Tanzi, E. Lucioni, S. Chaudhuri, L. Gori, A. Kumar, C. D'Errico, M. Inguscio, and G. Modugno, Phys. Rev. Lett. {\bf 111}, 115301 (2013).
% Transport of a Bose Gas in 1D Disordered Lattices at the Fluid-Insulator Transition %

%%% Superfluid in a disordered environment:theory %%%
\bibitem{Paul} T. Paul, P. Schlagheck, P. Leboeuf, and N. Pavloff, Phys. Rev. Lett. {\bf 98}, 210602 (2007).
\bibitem{Albert} M. Albert, T. Paul, N. Pavloff, and P. Leboeuf, Phys. Rev. Lett. {\bf 100}, 250405 (2008); Phys. Rev. A {\bf 82}, 011602(R) (2010).

%%% Phase slip due to thermal or quatum fluctuation %%%
\bibitem{Polkovnikov} A. Polkovnikov, E. Altman, E. Demler, B. Halperin, and M.D. Lukin, Phys. Rev. A {\bf 71}, 063613 (2005).
% Decay of superfluid currents in a moving system of strongly interacting bosons %
\bibitem{Danshita13} I. Danshita, Phys. Rev. Lett. {\bf 111}, 025303 (2013).
% Universal Damping Behavior of Dipole Oscillations of One-Dimensional Ultracold Gases Induced by Quantum Phase Slips %

%%% The disordered Bose-Hubbard model: theory  %%%
\bibitem{Fisher} M.P.A Fisher, P.B. Weichman, G. Grinstein, and D.S. Fisher, Phys. Rev. B {\bf 40}, 546 (1989).
% Boson localization and the superfluid-insulator transition %
\bibitem{Singh-92} K.G. Singh and D.S. Rokhsar, Phys. Rev. B {\bf 46}, 3002 (1992).
% Real-space renormalization study of disordered interacting bosons %
\bibitem{Freericks} J.K. Freericks and H. Monien, Phys. Rev. B {\bf 53}, 2691 (1996).
% Strong-coupling expansions for the pure and disordered Bose-Hubbard model %
\bibitem{Buonsante} P. Buonsante, F. Massel, V. Penna, and A. Vezzani, Phys. Rev. A {\bf 79}, 013623  (2009).
% Gutzwiller approach to the Bose-Hubbard model with random local impurities %
\bibitem{Bissbort} U. Bissbort, R. Thomale, and W. Hofstetter, Phys. Rev. A {\bf 81}, 063643 (2010).
% Stochastic mean-field theory: Method and application to the disordered Bose-Hubbard model at finite temperature and speckle disorder %
\bibitem{Altman} E. Altman, Y. Kafri, A. Polkovnikov, and G. Refael, Phys. Rev. B {\bf 81}, 174528 (2010).
% Superfluid-insulator transition of disordered bosons in one dimension %
\bibitem{Soyler} S.G. S\"{o}yler, M. Kiselev, N.V. Prokof'ev, and B.V. Svistunov, Phys. Rev. Lett. {\bf 107}, 185301 (2011).
%  %
\bibitem{Ristivojevic} Z. Ristivojevic, A. Petkovic, P. Le Doussal, and T. Giamarchi, Phys. Rev. Lett. {\bf 109}, 026402 (2012).
% Phase Transition of Interacting Disordered Bosons in One Dimension %
\bibitem{Zuniga} J.P. Alvarez Zuniga, D.J. Luitz, G. Lemarie, and N. Laflorencie, Phys. Rev. Lett. {\bf 114}, 155301 (2015).
% Critical Properties of the Superfluid-Bose-Glass Transition in Two Dimensions %

%%% The disordered Bose-Hubbard model: experiment  %%%
\bibitem{White} M. White, M. Pasienski, D. McKay, S.Q. Zhou, D. Ceperley, and B. DeMarco, Phys. Rev. Lett. {\bf 102}, 055301 (2009).
% Strongly Interacting Bosons in a Disordered Optical Lattice %
\bibitem{Deissler} B. Deissler, M. Zaccanti, G. Roati, C. D'Errico, M. Fattori, M. Modugno, G. Modugno and M. Inguscio, Nat. Phys. {\bf 6}, 354 (2010).
% Delocalization of a disordered bosonic system by repulsive interactions %
\bibitem{D'Errico} C. D'Errico, E. Lucioni, L. Tanzi, L. Gori, G. Roux, I.P. McCulloch, T. Giamarchi, M. Inguscio, and G. Modugno, Phys. Rev. Lett. {\bf 113}, 095301 (2014).
% Observation of a Disordered Bosonic Insulator from Weak to Strong Interactions %

%%% Standard Bogoliubov theory %%%
\bibitem{Fetter} A.L. Fetter and J.D. Walecka, {\it Quantum Theory of Many Particle Systems} (McGraw-Hill, Boston, 1971).

%%% Bogoliubov excitation in disordered Bose gases %%%
\bibitem{Singh-94} K.G. Singh and D.S. Rokhsar, Phys. Rev. B {\bf 49}, 9013 (1994).
% Disordered bosons: Condensate and excitations %
\bibitem{Gaul} C. Gaul, and C.A. M\"{u}ller, Phys. Rev. A {\bf 83}, 063629 (2011).
% Bogoliubov excitations of disordered Bose-Einstein condensates %
\bibitem{Lugan} P. Lugan and L. Sanchez-Palencia, Phys. Rev. A {\bf 84}, 013612 (2011).
% Localization of Bogoliubov quasi-particles in interacting Bose gases with correlated disorder %
\bibitem{Fontanesi} L. Fontanesi, M. Wouters, and V. Savona, Phys. Rev. A {\bf 81}, 053603 (2010).
% Mean-field phase diagram of the one-dimensional Bose gas in a disorder potential %
\bibitem{Saliba} J. Saliba, P. Lugan, and V. Savona, Phys. Rev. A {\bf 90}, 031603 (2014).
% Superfluid-insulator transition of two-dimensional disordered Bose gases %

%%% Bogoliubov theory for disordered Boson %%%
\bibitem{Huang} K. Huang and H.F. Meng, Phys. Rev. Lett. {\bf 69}, 644 (1992).
\bibitem{Falco} G.M. Falco, A. Pelster, and R. Graham, Phys. Rev. A {\bf 75}, 063619 (2007).
\bibitem{Yukalov07-1} V.I. Yukalov and R. Graham, Phys. Rev. A {\bf 75}, 023619 (2007).
\bibitem{Yukalov07-2} V.I. Yukalov, E.P. Yukalova, K.V. Krutitsky, and R. Graham, Phys. Rev. A {\bf 76}, 053623 (2007).

%%% Hohenberg-Martin dilemma %%%
\bibitem{Griffin} A. Griffin, Phys. Rev. B 53, 9341 (1996).
% Conserving and gapless approximations for an inhomogeneous Bose gas at finite temperatures %

%%% Bogoliubov theory: representative statistical ensemble %%%
\bibitem{Yukalov06} V.I. Yukalov and E.P. Yukalova, Phys. Rev. A {\bf 74}, 063623 (2006).
% Bose-Einstein-condensed gases with arbitrary strong interactions %
\bibitem{Yukalov08} V.I. Yukalov, Ann. Phys. (N.Y.) {\bf 323}, 461 (2008).
% Representative statistical ensembles for Bose systems with broken gauge symmetry %

%%% Equivalence between the one-particle spectrum and the hydrodynamic spectrum %%%
\bibitem{Gavoret} J. Gavoret and P. Nozi\`eres, Ann. Phys. (N.Y.) {\bf 28}, 349 (1964).
% Structure of the perturbation expansion for the bose liquid at zero temperature %

%%% Relation between the superfluid-Bose-glass transition point and the critical velocity %%%
\bibitem{Buonsante15} P. Buonsante, L. Pezze, and A. Smerzi, Phys. Rev. A {\bf 91}, 031601(R) (2015).
% Interacting bosons in a disordered lattice: Dynamical characterization of the quantum phase diagram %

%%% One-dimensional Bose gas %%%
\bibitem{Cherny} A.Y. Cherny, J. Caux, J. Brand, Front. Phys., {\bf 7}, 54 (2012).
% Theory of superfluidity and drag force in the one-dimensional Bose gas %

%%% Critical velocity at finite temperature %%%
\bibitem{Navez} P. Navez and R. Graham, Phys. Rev. A {\bf 73}, 043612 (2006).
% Subsonic critical velocity of a Bose-Einstein condensate at finite temperature %

\end{thebibliography}
\end{document}